\documentclass{ws-ijmpa}

\begin{document}

\markboth{Agnieszka Furman, Leonard Le\'sniak}
{Is the $a_0(980)$ resonance a $K\overline{K}$ bound state?}

%
\catchline{}{}{}{}{}
%

\title{
IS THE $a_0(980)$ RESONANCE A $K\overline{K}$ BOUND STATE?}

\author{\footnotesize AGNIESZKA FURMAN and 
        LEONARD LE\'SNIAK}

\address{The Henryk Niewodnicza\'nski Institute of Nuclear Physics\\
Polish Academy of Sciences\\
31-342 Krak\'ow, Poland}

\maketitle


\begin{abstract}
We have analysed properties of two resonances $a_0(980)$ and 
$a_0(1450)$ using the $\pi\eta$ and $K\overline{K}$ coupled channel 
model. 
Although the forces in the scalar-isovector $K\overline{K}$ channel are 
attractive the $a_0(980)$ resonance cannot be interpreted as a 
kaon-antikaon bound state within our model.

\end{abstract}

\section{Introduction}

Scalars are very controversial mesonic states. 
Their internal structure is not yet understood. 
There are different models in which the scalars are treated as simple 
$q\overline{q}$ mesons, $qq\overline{q}\overline{q}$ states, 
$K\overline{K}$ bound states or mixed states including glueballs. 
Two scalar-isovector resonances $a_0(980)$ and $a_0(1450)$ have been 
observed experimentally.
The $a_0(980)$ mass is close to the $K\overline{K}$ threshold so one 
should check whether the $a_0(980)$ is a $K\overline{K}$ quasi-bound 
state.

We study the properties of the $a_0$'s within the $\pi\eta$ and 
$K\overline{K}$ coupled channel model described in Ref.~\refcite{LL96}. 
It is based on the separable interactions between mesons: 
\begin{equation}
 <p|V_{ij}|q>=\lambda_{ij}f_i(p)f_j(q),\qquad 
    i,\,j=1~({\rm for~}\pi\eta){\rm~or~}2~({\rm for~}K\overline{K}),
\end{equation}
where $p$, $q$ are the c. m. momenta, $\lambda_{ij}$ are the coupling 
constants, $f_i(p)=1/(p^2+\beta_i^2)$ are the form factors and 
$\beta_i$ are the range parameters. 
We fix four parameters $\lambda_{11}$, $\lambda_{22}$, 
$\lambda_{12}$ and $\beta_2$ by choosing the $S$-matrix poles 
related to both $a_0(980)$ and $a_0(1450)$ resonances and  the fifth 
parameter $\beta_1$ by comparing the $K\overline{K}/\pi\eta$ branching 
ratio for the $a_0(980)$ with the experimental value.\cite{CB98} 
Below we present some of our results. 
More of them can be found in Refs.~\refcite{my02,my03}.

\section{Results}

\begin{figure}[h]
\centerline{\includegraphics*[height=4.4cm]{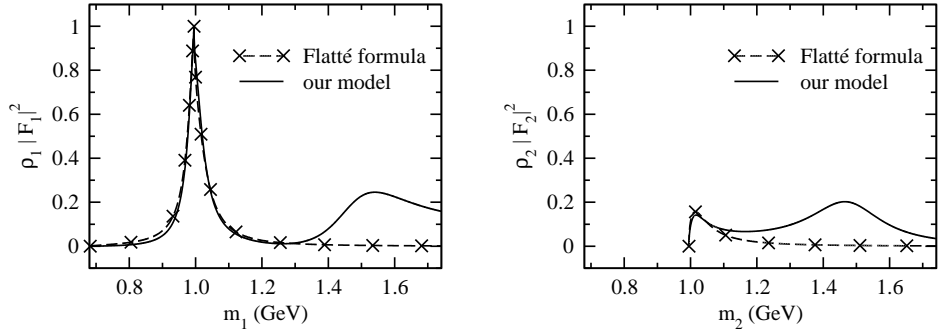}}
\vspace*{8pt}
\caption{Effective mass distributions in the $\pi\eta$ and 
         $K\overline{K}$ 
         channels}\label{fig:distrib}
\end{figure}

In Fig.~\ref{fig:distrib} the effective mass distributions obtained in 
the Flatt\'e parametrisation of the Crystal Barrel Collaboration 
data \cite{CB98} and in the coupled channel model \cite{my02} are 
compared. 
The phase space factors are defined as $\rho_i=2k_i/m_i$ and the 
production amplitudes are given by 
\begin{equation}
F_i=\frac{Ng_i}{m_0^2-m_i^2-i(\rho_1g_1^2+\rho_2g_2^2)},
\end{equation} 
where $m_i$ are the effective masses, $k_i$ are the channel momenta, 
$g_i$ are the coupling constants, $m_0$ is the mass parameter and $N$ 
is the normalization constant. 
The production amplitudes are related to the elastic and the transition 
cross sections by: 
\begin{equation}
\rho_1|F_1|^2=\sigma_{\pi\eta}^{el}k_1m_1 
\end{equation} and 
\begin{equation}
\rho_2|F_2|^2=\sigma_{\pi\eta \to K\overline{K}\,}k_1m_1. 
\end{equation}

The Flatt\'e formula describes only a range of the mass distributions 
near the $a_0(980)$ resonance while our coupled channel model describes 
both the $a_0(980)$ and the $a_0(1450)$ resonances. 
For example, we can predict the $K\overline{K}/\pi\eta$ branching ratio 
in the $a_0(1450)$ mass range. 
We have calculated it in two mass ranges. 
For a typical range, between $M_1=M-\Gamma/2$ and $M_2=M+\Gamma/2$, 
where $M=1474$~MeV and $\Gamma=265$~MeV one obtains a value of $0.98$. 
The branching ratio is equal to $0.78$ if $M_1=1300$~MeV and 
$M_2=1471$~MeV. 
Our predictions are in a good agreement with the experimental value 
$0.88\pm0.23$ given by the Crystal Barrel Collaboration.\cite{CB98}

\begin{figure}[h]
\centerline{\psfig{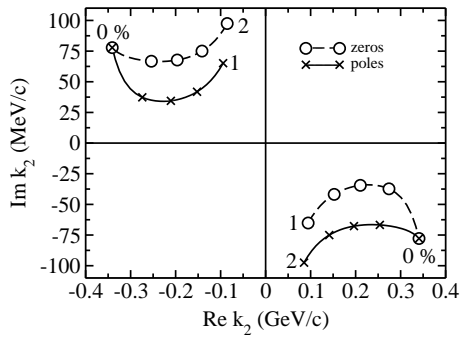}}
\vspace*{8pt}
\caption{Trajectories of the $S_{22}$ matrix element poles and 
         zeros related to the $a_0(980)$ resonance in the 
         $K\overline{K}$ complex momentum plane. 
         Crosses and circles indicate a reduction of the interchannel 
         coupling $\lambda_{12}^2$ by 25\%.}\label{fig:traj}
\end{figure}

Each resonance decaying in two channels is related to two poles of 
the $S$-matrix in the complex momentum planes. 
The $a_0(980)$ resonance lies close to the $K\overline{K}$ threshold.  
Both poles corresponding to this state are close to the physical 
region and influence strongly the $\pi\eta$ and $K\overline{K}$ 
amplitudes. 
Having fixed all the model parameters we have obtained an attractive 
force in the $K\overline{K}$ channel. 
Now the question is whether a strength of the $K\overline{K}$ 
force is sufficient to create a kaon-antikaon bound state. 
A pole position in the $K\overline{K}$ {\it uncoupled} channel gives us 
some information about the nature of the $a_0(980)$ state. 
If a bound state exist then it is connected with a pole on the positive 
part of the imaginary axis in the complex momentum plane. 
To arrive to a conclusion we have gradually {\it reduced} the 
interchannel coupling from its value down to zero and studied the 
pole trajectories. 
In Fig.~\ref{fig:traj} the trajectories of two $a_0(980)$ poles and 
two zeros are shown. 
In the uncoupled case the pole~$1$ meets the zero related to the 
pole~$2$ and is cancelled by it. 
In the same way the pole~$2$ is cancelled by the zero related to the 
pole~$1$. 
It means that in this limit both poles disappear from the 
$K\overline{K}$ channel so the $a_0(980)$ resonance cannot be 
interpreted as a $K\overline{K}$ bound state. 

\section{Conclusions}

We have constructed the $\pi \eta$ and $K\overline{K}$ coupled channel 
model. 
This five-parameter unitary model allows one to describe simultaneously 
both the $a_0(980)$ and $a_0(1450)$ states. 
All the parameters have been fitted to experimental values of the 
masses and widths of both $a_0$'s and to the $K\overline{K}/\pi \eta$ 
branching ratio measured near the $K\overline{K}$ threshold. 
The production amplitudes near the $a_0(980)$ resonance and 
the $K\overline{K}/\pi \eta$ branching ratio for the $a_0(1450)$ 
resonance, predicted by us, are in good agreement with the Crystal 
Barrel Collaboration results. 
Within our model and taking into account the existing experimental data 
the $a_0(980)$ resonance cannot be interpreted as a $K\overline{K}$ bound 
state.

\end{document}